\documentclass[twocolumn,aps,prb]{revtex4}

\usepackage{epsfig,graphicx,times}

\usepackage{amssymb}
\usepackage{amsmath}
\usepackage{graphicx}
\usepackage{dcolumn}
\usepackage{bm}

\begin{document}

\title{Measurement of Magnetization Dynamics in Single-Molecule Magnets Induced by Pulsed Millimeter-Wave Radiation}

\author{M. Bal}
\author{Jonathan R. Friedman$^{*}$}
\affiliation{\mbox{Department of Physics, Amherst College,
Amherst, Massachusetts 01002-5000, USA}}

\author{M. T. Tuominen}
\affiliation{\mbox{ Department of Physics, University of
Massachusetts, Amherst, Massachusetts 01003, USA}}

\author{E. M. Rumberger}
\author{D. N. Hendrickson}
\affiliation{\mbox{ Department of Chemistry and Biochemistry,
University of California at San Diego, La Jolla, California 92093,
USA}}

\date{\today}

\begin{abstract}
\noindent We describe an experiment aimed at measuring the spin
dynamics of the Fe$_8$ single-molecule magnet in the presence of
pulsed microwave radiation. In earlier work, heating was observed
after a $0.2$-ms pulse of intense radiation, indicating that the
spin system and the lattice were out of thermal equilibrium at
millisecond time scale [Bal et al., Europhys.~Lett.~\textbf{71}, 110
(2005)]. In the current work, an inductive pick-up loop is used to
probe the photon-induced magnetization dynamics between only two
levels of the spin system at much shorter time scales (from ns to
$\mu$s). The relaxation time for the magnetization, induced by a
pulse of radiation, is found to be on the order of 10 $\mu$s.
\end{abstract}

\maketitle

Single-molecule magnets are unique systems that lie at the border
between the classical and quantum worlds. Like classical magnets,
they are bistable and exhibit hysteresis at low
temperatures.\cite{18, 10} They also exhibit fascinating quantum
mechanical properties, such as tunneling between ``up'' and
``down'' orientations,\cite{33, 81, 91} and interference between
tunneling paths.\cite{162} Furthermore, their potential use in
quantum computation has been proposed,\cite{298} and experiments
with millimeter-wave radiation have demonstrated an enhancement in
the relaxation rate for magnetization reversal,\cite{335, 283} as
well as induced changes in the equilibrium
magnetization.\cite{345, 343, 339, 337, 290}

Recent work has focused on measuring the spin relaxation time $T_1$
and the decoherence time $T_2$.\cite{339, 345, 346, 352}
Determination of these parameters as well as exploration of coherent
spin dynamics are important in order to see if single-molecule
magnets can be employed as qubits.\cite{298} In previous work, the
speed of the micro-Hall-bar detectors used prevented the measurement
of magnetization relaxation at time scales shorter than a few
microseconds.\cite{345} In this paper, we use a thin-film inductive
pick-up loop as a fast flux detector.  This method, however, has the
drawback of reduced sensitivity when compared to the Hall sensor.

We study the single-molecule magnet
Fe$_8$O$_2$(OH)$_{12}$(tacn)$_6$ (hereafter called Fe$_8$), which
is composed of eight magnetic Fe(III) ions strongly coupled
together to form a single spin-10 system and is described by the
effective Hamiltonian

\begin{equation}
{\cal H} =  - DS_z^2  + E(S_x^2-S_y^2)+C(S_+^4+S_-^4)- g\mu _B
\vec{S} \cdot  \vec{H}, \label{Ham}
\end{equation}

\noindent where the anisotropy constants $D$, $E$, and $C$ are
0.292 K, 0.046 K, and $-2.9$ $\times$ 10$^{-5}$K, respectively,
and $g$ = 2.\cite{162, 188, 184, 350} The first (and largest) term
causes the spin to prefer to lie along or opposite the z axis,
resulting in a double-well potential, and making the energy levels
approximately the eigenstates of $S_z$. The second and third terms
break the rotational symmetry of the Hamiltonian and result in
tunneling between the otherwise unperturbed states. Reversal of
the magnetization from one easy-axis direction to another is
impeded by a $\sim 25$ K barrier.\cite{91}

\begin{figure}[htb]
\centering
\includegraphics[width=80mm]{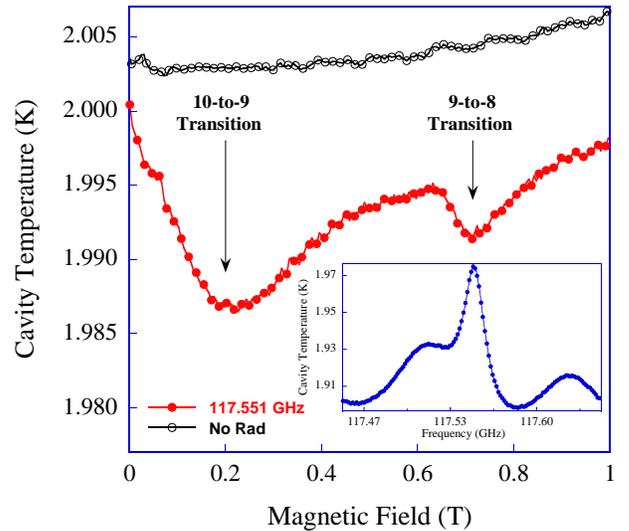} 
\caption{(Color online) Cavity temperature as a
function of magnetic field,  both in the presence and absence of
continuous-wave radiation. Transitions are observed from the m = 10
level to the m = 9 level, as well as from m = 9 to m = 8, as
indicated. The resonant frequency of the cavity is determined by
monitoring the cavity temperature, as shown in the inset. When on resonance
the temperature rises sharply because radiation enters the cavity.  The
variation of the cavity temperature away from the resonance is due
to the variation of the output power of the microwave source with
frequency.
\label{spectroscopy} }
\end{figure}

As a radiation source we use either a solid-state multiplier fed by
a frequency synthesizer or a high-power backward-wave oscillator
(BWO). The radiation propagates in a rectangular stainless-steel WR-10 waveguide
with gold plating on the inner surfaces. Our resonator is a cylindrical cavity
made of oxygen-free copper with dimensions of 3.17 mm $\times$ 4.62
mm (diameter $\times$ length). Some resonances,
such as the $TE_{0,1,1}$ mode, of a cylindrical cavity do not
require an ac current to flow between the end plates and the walls,
resulting in very high quality factors (on the order of $\sim10^4$
in the present study).\cite{301} Finally, a thin-film pick-up loop
on a silicon substrate serves as our detector for measuring
changes in magnetic field originating from a single crystal of
Fe$_8$  placed inside the cavity. The pick-up
loop is fabricated using standard electron-beam lithography,
followed by metal deposition (thermally evaporated Cu film) and
lift-off. A rectangular loop ($65 \times 205$ $\mu$m$^2$) is defined
by 5-$\mu$m-wide and $200$-nm-thick Cu lines.

Radiation is coupled to the cavity through a small hole
in the top plate of the cavity. A $25$-$\mu$m-thick gold foil forms
the bottom plate. A $\sim0.5$-mm-diameter hole is made in the Au
foil and the pickup loop is placed just beneath this hole.  The hole
allows the fast magnetization signal from the sample to couple to
the loop without attenuation and also permits the sample to be
positioned as close as possible to the pick-up loop. In some experiments, a thin piece of
Mylar foil coated with $50$ nm of amorphous copper is placed between
the sample and loop. The copper film is much thicker than the skin
depth of the radiation, allowing the foil to form a good
electromagnetic closure for the cavity at microwave frequencies. At
the same time, the film is thinner than the skin depth at the time
scales of the sample magnetization signals, allowing them to couple
through the film without significant attenuation. The external magnetic field
is applied along the axis of the cylindrical cavity and
perpendicular to the microwave field at the sample location.

The Fe$_8$ sample is mounted such that its crystallographic b-axis
is parallel to the substrate surface of the pick-up loop. The loop
dimensions were chosen to match the typical size of a single
crystal. Both of these considerations allow for maximum flux
coupling.  However, the sample mounting technique results in a large angle (typically around
$35^\circ$ to $45^\circ$) between the easy-axis of the magnetization and
the applied magnetic field. The induced emf in our pick-up loop is
directly proportional to the rate of change in the magnetization
(dM/dt) of the Fe$_8$ sample. We also did experiments with the
sample's a-axis parallel to the external field (in order to reduce
the angle between the easy axis and the field to $\sim$15$^\circ$),
but were unable to measure a signal because the flux coupling was
too weak.

We determined the resonant frequency of the cavity by monitoring the
cavity temperature while sweeping the frequency of the radiation.
The finite conductivity of the cavity walls results in ohmic heating
when on resonance and is registered by our thermometer (mounted
outside the cavity), as shown in the inset of Figure
\ref{spectroscopy}. This resonance, at 117.551 GHz, corresponds to the
$TE_{0,1,1}$ mode of the cylindrical cavity and has a quality factor
of $\sim9800$.

\begin{figure}[htb]
\centering
\includegraphics[width=79mm]{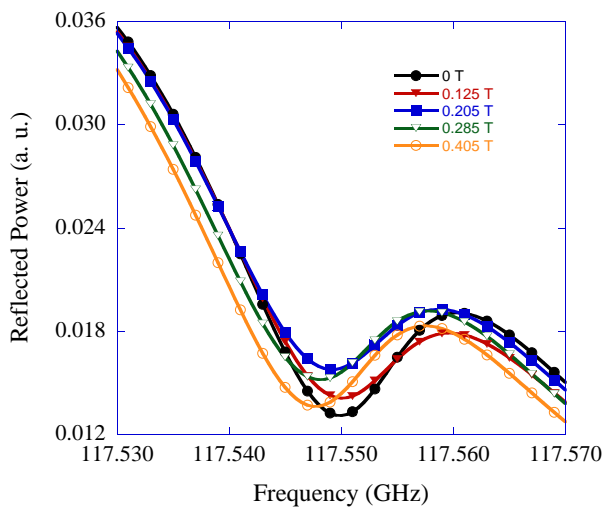} 
\caption{(Color online) Reflected power from the cavity is
measured as a function frequency and is plotted for several
constant magnetic field as indicated on the plot. \label{cavres} }
\end{figure}

The pick-up loop is not a suitable detector of slowly varying
magnetic signals and therefore is not useful to perform
spectroscopic studies. Measuring the cavity temperature, on the
other hand, as a function of the magnetic field allowed us to
perform crude spectroscopic measurements. In Figure
\ref{spectroscopy}, the cavity temperature is shown as a
function of
magnetic field and clearly changes when the resonance condition for
the Fe$_8$ sample is satisfied (10-to-9 and 9-to-8 transitions are
shown); the measurement taken in the absence of radiation is shown
for comparison. It should be noted, however, that the cavity
temperature in Fig.\ref{spectroscopy} decreases when the sample is
on resonance with the radiation, in contrast to the heating observed
in our previous experiment.\cite{345} This unexpected behavior
required us to study the properties of the cavity in greater detail.
Figure \ref{cavres} shows the microwave power reflected from the
cavity, using a directional coupler and diode detector, as a
function of frequency at several constant magnetic fields. The
microwave power reflected from the cavity changes as we field tune
the Fe$_8$ sample through the 10-to-9 transition. In other words,
the power absorption by the cavity is minimized when the resonance
condition is met. A frequency shift of a few MHz is also present but
does not seem to depend on whether the sample is on resonance with
the radiation. The decrease in cavity temperature can be explained
as a consequence of a decrease in the absorbed microwave power by
the cavity, which in turn is caused by the sample coming to resonance with the cavity.
These observations suggest that tuning the sample to resonance with
the cavity adds a dispersive element to the cavity and thereby
changes the impedance matching condition between the waveguide and
the cavity. Similar cavity-sample interactions form the basis of
electron-spin-resonance spectrometers in which a resonant cavity can
be operated either in the absorption mode (quality factor change) or
in the dispersion mode (frequency change).\cite{348}

\begin{figure}[htb]
\centering
\includegraphics[width=80mm]{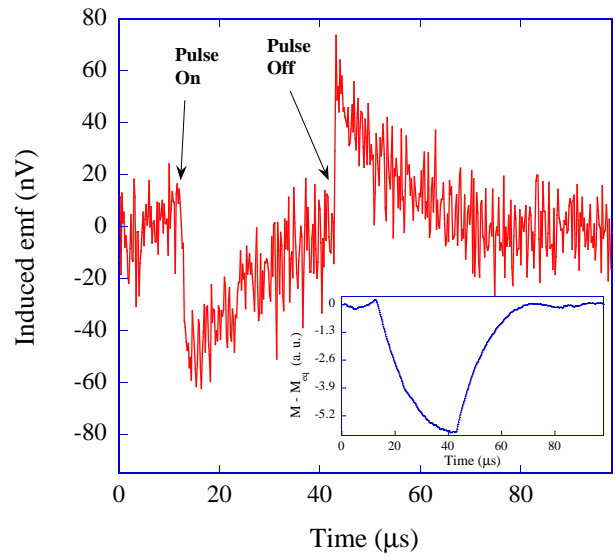} 
\caption{(Color online) The induced emf (proportional to dM/dt) in
the pick-up loop is plotted as a function of time. The inset shows
the magnetization, calculated by numerically integrating the induced
emf, as a function of time. \label{indemf} }
\end{figure}

A PIN-diode switch is employed to pulse the microwave radiation. The
radiation frequency is set to the cavity resonance as described
above. The field is set $0.19$ T so that the transition from the ground
to the first excited state of the Fe$_8$ sample is resonant with
the radiation (Fig. \ref{spectroscopy}). A 30-$\mu$s radiation pulse is applied and repeated
every 10 ms at a temperature of $1.8$ K. A fast digitizing
oscilloscope is used to capture the induced emf from the pick-up
loop after passing through a fast, low-noise preamp. Figure
\ref{indemf} shows the induced emf as a function of time; the data
contains the average of 3.84 $\times$ 10$^5$ individual traces to
increase the signal-to-noise ratio. The inset shows the sample
magnetization as a function of time, obtained by numerically
integrating the data in the main part of the figure. The induced emf
(and thereby dM/dt) exhibits sharp jumps when the radiation is
turned on and off. The sharp jumps indicate that the radiation
immediately causes the magnetization to start decreasing and that it
begins to increase as soon as the radiation is turned off, as shown
in the inset. The rate of magnetization change decays to zero in a
time of roughly 10 $\mu$s. After such a time, the magnetization has
achieved a nearly steady state.

We tentatively interpret these results in terms of the spin dynamics
of the two levels involved in the radiative transition. Turning on
the radiation initiates photon-induced transitions from m = 10 to
9,\cite{500} resulting in a decrease in the magnetization. After a
steady state is reached, the m = 9 state has a larger population than
under thermal equilibrium conditions and the excess population
decays back to the ground state once the radiation is turned off.
The $\sim$10-$\mu$s relaxation time is surprisingly long for
spin-phonon interactions, which presumably determine the time scale
of the observed decays. Further study of the spin dynamics in the
presence of pulsed radiation is needed to understand the origin of
this long relaxation time.

In conclusion, we have built a cryogenic millimeter-wave probe to
measure the fast magnetization dynamics of
single-molecule magnets in the presence of pulsed, resonant
radiation. A cylindrical cavity with a high quality factor and a
thin-film inductive pick-up loop are used to study the time
evolution of the magnetization of Fe$_8$ during and after a single
pulse of millimeter-wave radiation. We find an unusually long
relaxation time of $\sim$10-$\mu$s.  A poor signal-to-noise ratio
requires the averaging of a large number of traces, which currently
demands extremely long experimental run times. Modifications to the
experimental set-up to improve the signal-to-noise ratio are in
progress.

We thank M. P. Sarachik, Y. Suzuki, D. DeMille, J. Tu, D. Hall, L.
Hunter, and K. Mertes for useful conversations.  We also thank D.
Krause and G. Gallo for their technical contributions to this
study. We are also grateful to M. Foss-Feig, E. da Silva Neto, and
J. Rasowsky for their help in acquiring and analyzing some of
the data. Support for this work was provided by the US National
Science Foundation, the Research Corporation, the Alfred P. Sloan
Foundation, and the Center of Excellence of the Israel Science
Foundation.


\begin{thebibliography}{10}
\bibitem{500}We note that the energy eigenstates are not pure $m$ states
because of the large angle between the external magnetic field and
the easy axis of Fe$_8$ crystal, as well as because of transverse
anisotropy terms in the Hamiltonian.  However, the magnetic field
used in this work (0.19 T) is small relative to the anisotropy fields
and therefore does not result in
any significant mixing of the eigenstates. For example, we calculate the
difference between the mixed eigenstate and the pure m = 10 (m = 9)
state to be 0.37 $\%$ (1.25 $\%$) for an angle of 45$^\circ$.
\end{thebibliography}
\end{document}